\documentclass[aps,superscriptaddress,pra,twocolumn,secnumarabic,nobalancelastpage,amsmath,amssymb,nofootinbib]{revtex4}

\usepackage{graphics}      
\usepackage[pdftex]{graphicx,color}
\usepackage{longtable}     
\usepackage{url}           
\usepackage{bm}            

\usepackage{amsmath}
\usepackage{amssymb}

\newcommand\bra[1]{\left<#1\right|}
\newcommand\ket[1]{\left|#1\right>}
\newcommand\bracket[2]{\left<#1|#2\right>}
\renewcommand\Re[1]{\text{Re}\left[#1\right]}
\renewcommand\Im[1]{\text{Im}\left[#1\right]}

\begin{document}

\title{Recoil Free Scattering From a Free Gas}
\author{Scott N. Sanders}
\affiliation{Massachusetts Institute of Technology, Cambridge,
  Massachusetts 02139}
\email{ssanders@post.harvard.edu}
\author{Florian Mintert}
\affiliation{Albert-Ludwigs-Universit\"{a}t Freiburg,
  Hermann-Herder-Str. 3, 79104 Freiburg}
\affiliation{Harvard University, Cambridge,
  Massachusetts 02138}
\author{Eric J. Heller}
\affiliation{Harvard University, Cambridge,
  Massachusetts 02138}
\date{\today}

\begin{abstract}
  We present a treatment of decoherence in an atom due to scattering
  from a gas of free particles. We show that there is a recoil
  free scattering process that leaves both the atom and the gas in an
  unchanged state, but allows for the acquisition of a phase shift
  that remains in the free space limit. This is essential to
  understanding decoherence in a separated arm atom interferometer,
  where a gas of atoms forms a refractive medium for a matter
  wave. Our work clarifies the extent to which scattering of a free
  particle acts as a which-way measurement.
\end{abstract}

\maketitle

\section{Introduction}
Can a propagating atom run the gauntlet through a gas of free atoms,
interacting with all of them at long range, and still remain
coherent, leaving the quantum state of every gas atom unchanged?  The
answer is yes, much of the time, depending on gas density, propagation
distance, atom-atom collision cross sections, etc. (the answer is no,
however, if the force is Coulombic). If this were not the case, the
measurement in \cite{Pritchard1995-1} of the refractive index of a gas
of atoms for an atomic matter wave would not have worked. These
measurements took place in a separated arm atom interferometer, where
one arm intersected a gas cell, filled with other atoms. Any collision
that had disturbed the state of an atom in the gas cell would have
been a which-way measurement that reduced the interference fringe
contrast of the interferometer.

A simple estimate of the coherent part of the propagating wave can be
made using known atom-atom elastic quantum cross sections, computing
the chance of avoiding a collision in the usual way. However, this
leaves another question unanswered: if there is a large survival
rate, avoiding any collisions, can the phase shift acquired by the coherent
atom wavefunction be large compared to $\pi$?.

It is well known that matter can act as a coherent, refracting medium
for matter waves, as for example in the propagation of neutrons
through condensed matter. In passing through a solid, neutrons may
acquire large phase shifts relative to the vacuum and emerge
coherently, without changing the quantum state of the material at all.
However solids are rather rigid compared to a low density gas; to wit,
consider neutron diffraction from a crystal; the elastic diffractive
spots are {\it prima facia} evidence of coherent scattering from the
crystal. It is thus surprising perhaps that atoms passing through
gaseous matter can also acquire large phase shifts without leaving a
trace of their passing, since gas atoms are so easily perturbed.  Our
analysis is in the context of a separated arm atom interferometer,
with one arm intersecting a cell containing a fixed density of gaseous
atoms (Fig.~\ref{fig:decoherence}). As we will see, a low density gas
is completely intolerant of any momentum transfer; momentum transfer
will always lead to decoherence.

The separated arm atom interferometer exploits superposition by
splitting the spatial wave function of an atom into two wavepackets
that can be made to travel along separate paths and experience
different interactions. It takes advantage precisely of the ability of
quantum systems to exist in superposition states. Such an apparatus
is, consequently, a highly sensitive detector of decoherence. In the
language of decoherence theory, the atom passing through the
interferometer is the system, and the free gas it interacts with is
the ``environment''. ``Leaving a trace of passing'' in the gas is a
which-way detection that causes decoherence and loss of fringe
contrast when the arms of the interferometer are recombined.

The interference fringe contrast is defined precisely in
\cite{Cronin2005} as $C=(I_{max}-I_{min})/(I_{max}+I_{min})$, where
$I$, the count rate in their detector, is proportional to the
probability density of detecting a projectile at a particular position
on the screen in Fig.~\ref{fig:decoherence}. The contrast of the
interference pattern is reduced due to correlation with the
environment \cite{Walls1993}. This point of view is equivalent to a
``which-way'' detection, in which the state of the environment records
partial or complete path information.

The relative phase acquired along the different paths of the
interferometer is observed as a shift in the interference pattern that
forms when the wave packets
recombine~\cite{Pritchard1997,Berman}. Previous experiments have
measured the interference pattern due to the presence of a free gas in
only one arm of such an interferometer \cite{Pritchard1995-1}. The
other arm was physically separated from the gas, and did not interact
with it. The experiments showed that, like light passing through
glass, a matter wave passing through a dilute gas experiences a phase
shift, with the dilute gas acting as a medium with an index of
refraction for matter waves.

When produced by propagation through a free gas, the phase shift of
the interference fringes is a probe of the atom-atom interactions, and
was the focus of much theoretical
work~\cite{Pritchard1995-1,Dalgarno1996,Dalgarno2001}. These
treatments build upon the multiple scattering theory derived in
\cite{Lax1951}, and they neglect the possibility of recoil of the
background atoms. Only the projectile is treated explicitly quantum
mechanically. The background gas creates a background potential, and
decoherence is obtained by averaging the resulting scattered
projectile wavefunction over different realizations of the
potential. Our interest here is understanding why spatial coherence is
able to persist at all, so our treatment will explicitly account for
which-way information in the state of the background atoms.
 
Experiments have also been performed to measure the amount of decoherence
experienced by an atom due to the scattering of photons from a laser
\cite{Pritchard1995-2,Pritchard2001} and to the scattering of atoms in
a free background gas \cite{Zeilinger2003,Cronin2005}. The decoherence
is observed as a loss of contrast in the interference patterns formed.

The theoretical foundation of the analysis used to understand these
experiments postulates that scattering events can be described as an
instantaneous modification of the system-environment density matrix,
$\rho_i \rightarrow \rho_f = T \rho_i T^{\dagger}$
\cite{Tegmark1993}. The changes to the density matrix due to these
scattering events may be explicitly added to the usual Heisenberg
equation of motion. The additional term gives rise to decoherence of
the system when the degrees of freedom of the environment are traced
over.

The physical mechanism by which this process occurs, however, remains
hidden in the {\it ad hoc} addition to the purely coherent evolution
of the density matrix. Ideally, one would like an explanation of the
surprising lack of sensitivity of a free particle as a which-way
detector based on microscopic scattering theory. The effect on a
quantum particle due to a gas environment, treated as a Markovian
reservoir in which only two-body scattering is considered, has been
treated in a very general way by \cite{Hornberger2006}.

Our objective is rather different. We wish to show the origin of the
phase shift on an atom wave function due to scattering from other
atoms. We also seek an explanation of the surprising lack of
sensitivity of a free particle as a which-way detector based on
microscopic multiple scattering theory. We will, therefore, calculate
the reduction in interference fringe contrast due to the presence of a
free gas interacting with only one arm of a separated arm atom
interferometer. Our derivation shows how these processes emerge
directly from microscopic quantum mechanical scattering and avoids the
\textit{ad hoc} modification of the Heisenberg equation of motion and
the introduction of an average wave function. In fact, the coherent
wave introduced in \cite{Lax1951} {\it emerges} directly from our
calculations, providing a justification for its
use.

\section{Theoretical Perspective}

Standard scattering theory suggests a na\"{i}ve argument that little
or no coherence should remain after an atom passes through a column of
gas. The usual expression for scattering in free space (in the center
of mass frame) gives the wavefunction for the scattered atom as
\cite{Sakurai1994-1}
\begin{equation}
\label{eq:scattnormal}
\psi(\vec{r}) = \frac{1}{(2 \pi)^{3/2}} \left( e^{i \vec{k} \cdot \vec{r}} + f(\theta, \phi) \frac{e^{i k r}}{r} \right).
\end{equation}
The first term on the right-hand side of this equation is the
unscattered incident wave, which preserves coherence but has no phase
shift. A phase shifted, coherent contribution cannot arise from this
term. The second term on the right-hand side is the scattered wave,
which corresponds to a momentum conserving recoil of the target gas
atom, except when scattering into the exactly forward
direction. Scattering into the infinitesimal solid angle around the
forward direction occurs with zero probability. Any finite recoil
changes the state of a free target gas atom and constitutes a
which-way measurement that ought to eliminate the possibility of
observing interference. Only an infinitesimal fraction of the incident
beam would interact with the free gas atom and remain coherent. The
rest is either not scattered at all or decoheres completely.

Nonetheless, the experimental results~\cite{Pritchard1995-1} clearly
demonstrate that atoms in the beam do interact with the
background gas coherently because the phase shift that results from
the interaction is observable in the interference pattern. The beam
atoms are able to ``scatter'' off of the free gas atoms and acquire a
phase shift, without touching the free gas and changing its quantum
state at all.

A better approach to understanding the phase shift and the decoherence
is to enclose the target gas in a box, confining it in three
dimensions. We may then treat the interactions between a projectile
and a gas of confined particles. The projectile can be assumed to be
unaffected by the walls of the box through which it passes, as we will
eventually remove this artifice. The benefit of the box is immediate
-- there can be a finite amplitude associated with exactly ``forward''
scattering, in which the quantum state of the projectile and the
target are unaffected by the interaction. A key point is that {\it
  this coherent amplitude automatically comes with a nonvanishing
  phase shift}. The argument of the complex amplitude gives rise to a
phase shift, and its magnitude squared gives the probability of not
disturbing the environment in any way, and thus leaving the system
coherent. This result differs from free space scattering because there
will in general be a finite flux of system atoms that acquire a phase
shift and remain coherent. The coherent phase shift due to a single
target atom will approach zero as the cross sectional area of the
confining box is enlarged. This recalls the conundrum of
(\ref{eq:scattnormal}); however, the phase shift does not vanish, even
as the box is enlarged, if the column density of the gas remains
constant. It is a crucial task here to consider this limit carefully.

We will solve the problem of scattering of a beam atom, the
``system'', or projectile, from a gas of atoms, the
``environment''. The latter are confined to a three-dimensional box. The
beam atom itself will be confined to a waveguide that overlaps the
gas cell (Fig.~\ref{fig:waveguide}). In this way, the transverse modes of the beam atom
eigenstates are discrete, as are the modes in all three directions of
the gas atoms. We will assume that the beam atom does not feel the
confining wall that defines the length of the gas cell. We can then
study the interactions that lead to phase shifts of the beam atom
without changing the discrete state of the gas atoms. It is precisely
this recoiless interaction that gives rise to the coherent wave.

The imposition of a cell and a waveguide are reasonable in the context
of the experiments \cite{Pritchard1995-1}, where the gas was in fact
confined to a cell. The cell was macroscopically large, however, so
our results must not depend on the size of our cell. The relevant
experimental parameter is the column density of the gas. When we
consider the limit of large cell dimensions, we will choose the number
of gas atoms correspondingly, so that the column density remains
fixed. We will find that our results are independent of the dimensions
of the waveguide and gas cell and only depend on the column
density. In the limit where the cross section of the waveguide is very
large, our results explain the coherent interactions in free space
that cause a phase shift on the atom beam, while leaving the
background atoms completely untouched.

\section{Multiple Scattering Due to a Free Gas}

In a dilute gas, any scattering event which leads to recoil of a
target atom, placing it in an orthogonal state, leads also to complete
decoherence of the two-arm projectile density matrix. The orthogonal
target atom state constitutes which-way evidence and coherence cannot
persist. To calculate the total decoherence, we need only calculate
the amplitude of the many-particle state that remains unchanged by the
interactions, other than the acquisition of a phase shift. Over short
enough distances traveled by the projectile, we may neglect multiple
scattering altogether because the gas is dilute. If the projectile
survives the interactions over a short distance by remaining in the
initial state, then it is able to continue its journey toward the
detector and scatter downstream. The projectile can have many
sequential interactions with the gas atoms, so long as it remains in
its initial state after each scattering event. In this way, it can
accumulate a potentially large phase shift, even if the phase
accumulated by a single scattering event is small. After passing
through the entire cell, the amplitude of the initial state, which is
coherent with the other arm of the interferometer, will also
have been reduced due to scattering out of it.

\begin{figure}
\includegraphics{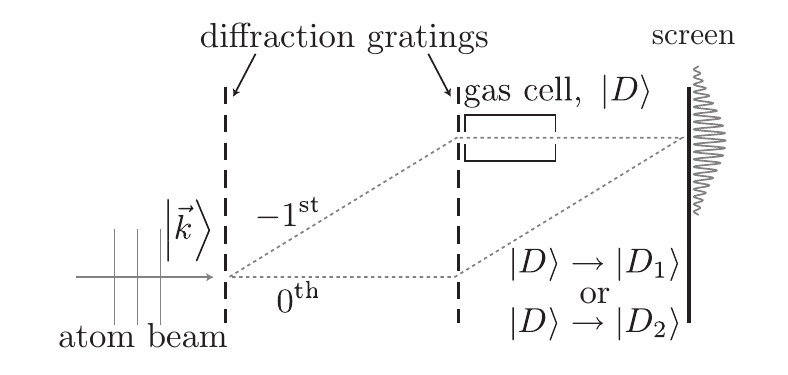}
\caption{\label{fig:decoherence}A Mach-Zender interferometer with the
  gas cell serving as a which-way detector. The atom beam is
  coherently split into the two arms of the interferometer by the
  leftmost diffraction grating. The initial state of the gas is
  $\ket{D}$, which evolves into $\ket{D_1}$ or $\ket{D_2}$ depending
  on whether an atom from the beam passes through the cell. An
  interference pattern forms on the screen at right where the arms
  overlap.}
\end{figure}

Fig.~\ref{fig:decoherence} shows the experimental configuration
we are considering.
The projectile passes coherently through the upper and lower arms of
the interferometer. A low pressure gas is present in the upper arm.

We model the upper arm as an overlapping waveguide and gas cell
(Fig.~\ref{fig:waveguide}). We discretize the transverse states of
the projectile atom by requiring that its wave function satisfy
periodic boundary conditions on the surface of the wave
guide. Similarly the states of the $N$ gas atoms are discretized by
requiring that they satisfy periodic boundary conditions on all
the surfaces of the gas cell. The projectile and target gas atoms are
otherwise free. The Hamiltonian describing this $N+1$ particle system,
in the absence of interactions, is $H_o$, with eigenstates $\ket{k,
  \vec{n}}$. The components of $\vec{n}$ are the $3N+2$ discrete
quantum numbers describing the transverse state of the projectile and
the states of the $N$ target atoms. $k$ is the initial longitudinal
wavenumber of the projectile.

For a dilute gas we neglect interactions between target atoms. The
interaction potential between the projectile and the targets is a sum
of binary terms. The projectile is labeled as the $0^{\text{th}}$
particle, and the targets will be labeled $1$ through $N$. The full
interaction potential, $V$, is then
\begin{equation}
\label{eq:interactionpotential}
V = \sum_{i=1}^{N} V_{0i}.
\end{equation}
$V_{0i}$ gives the potential between the projectile and the
$i^{\text{th}}$ target, and the full Hamiltonian is $H = H_o + V$. We
will take the projectile to be initially in an eigenstate of the
waveguide. Conservation of energy and momentum requires that if a
target remains in its initial state, then so must the projectile. 

The S-matrix connects the initial many body state, $\ket{k,
  \vec{n}_0}$, with the asymptotic output channel $\ket{\psi}$
\cite{Watson1964-1},
\begin{equation}
\ket{\psi} = S \ket{k, \vec{n}_0}.
\end{equation}
$\ket{\psi}$ is the many body state that emerges after interactions
between the projectile and the gas are complete. We will refer to the
diagonal element of the S-matrix that gives the $\ket{k, \vec{n}_0}$
component of $\ket{\psi}$ as $S_{0,0}$,
\begin{equation}
\label{eq:ScatteredWaveParts}
\ket{\psi} = S_{0,0} \ket{k, \vec{n}_0} + \text{orthogonal terms}.
\end{equation}
The first term on the right hand side of (\ref{eq:ScatteredWaveParts})
is the only part of $\ket{\psi}$ that interferes with the other arm of
the interferometer. The probability of finding the system plus
environment in this state is the probability that the system will
remain coherent and interfere with itself. The contrast of the
interference fringes will be reduced by the factor $\left| S_{0,0}
\right|$ \cite{Walls1993}. In order to calculate the amplitude of the
coherent state after interactions with the gas, we need to calculate
the $S_{0,0}$ matrix element. This task is facilitated by subdividing
the gas cell into thin slabs, and computing the contributions to
$S_{0,0}$ from each slab.

\subsection{Thin Slab Construction}

\begin{figure}
\includegraphics{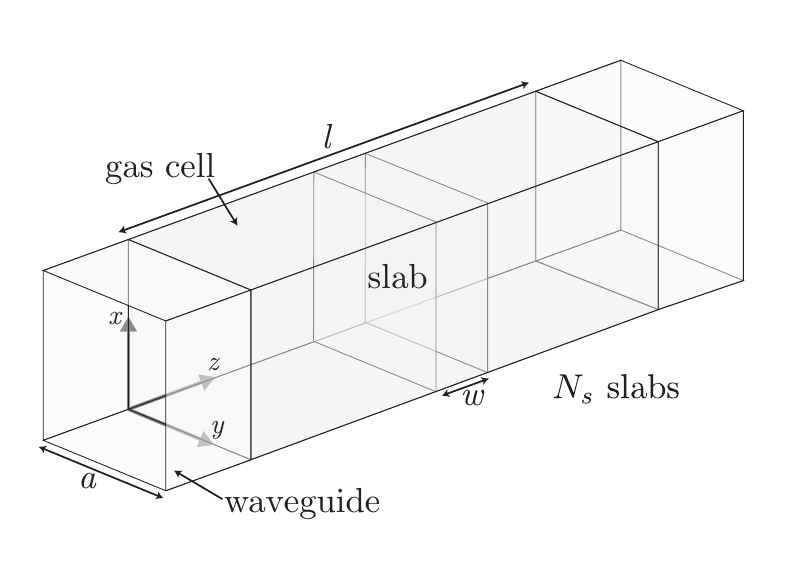}
\caption{\label{fig:waveguide} Details of the gas cell appearing in
  Fig.~\ref{fig:decoherence}. The cell has length, $l$, and is embedded in a
  waveguide. The $j^{\text{th}}$ slab, running from $z=(j-1) w$ to
  $z=j w$, is illustrated. The waveguide in which the projectile is
  confined is an infinite tube with a square cross section of
  dimensions, $a \times a$.}
\end{figure}

The volume of the gas cell can be thought of as the composition of
many adjacent, thin slabs, which are the regions of space formed by
the surface of the waveguide and two of its cross sections, placed a
distance $w$ apart, as in Fig.~\ref{fig:waveguide}. Imagine
subdividing the gas
cell into $N_s$ such regions, so that $l=N_s w$. If we number the
slabs, $j=1, 2, 3, \ldots$, beginning from the point of entry of the
projectile into the gas cell, then slab $j$ has the width and height
of the waveguide, and runs from $z= (j-1) w$ to $z=j w$.

The total interaction potential can be rewritten in terms of the
contribution from each slab,
\begin{eqnarray}
  V &=& \sum_{j=1}^{N_s} V^{(j)} = \sum_{j=1}^{N_s} \sum_{i=1}^{N} V_{0i}^{(j)}, \\
  V_{0i}^{(j)} &=& V_{0i} \ \theta(\hat{z}_i - (j-1) w) \, \theta(j w - \hat{z}_i).
\end{eqnarray}
$\hat{z_i}$ is the $z$-position operator for the $i^{\text{th}}$
target atom. $V^{(j)}$ picks out the contribution to the total
interaction potential due to a particular region of space. Summing
over these contributions, we obtain the original interaction
potential.

For each $V^{(j)}$, we will define a corresponding $S^{(j)}$, which is
the S-matrix due only to the interactions with the
$j^{\text{th}}$ slab. Beginning with the first slab, we can compute
the scattered state due only to that slab. If we then use that result
as the incident state to the subsequent slab (again removing all other
slabs), the state we will obtain after $N_s$ such iterations is
\begin{equation}
  \ket{\psi^{(N_s)}} = S^{(N_s)} S^{(N_s-1)} \cdots S^{(1)} \ket{k, \vec{n}_0}.
\end{equation}
$\ket{\psi^{(N_s)}}$ is different from $\ket{\psi}$ in general because
the wavefunction at earlier slabs is unaffected by subsequent slabs.
This excludes the possibility that the projectile could back scatter
but be recovered into the incident state by scattering a second time
from an earlier slab; however, for a dilute gas this process is
negligible, so we may safely approximate $\ket{\psi^{(N_s)}} \approx
\ket{\psi}$.

The decoherence, which causes the interference fringe contrast to be
reduced, is due to the reduced amplitude of the initial many-body
state. The phase shift that is measured as a spatial shift in the
observed interference fringes is given by the argument of that
amplitude. Equivalently, the magnitude of the overlap of the final
states of the free gas associated with each arm gives the decoherence
and the argument of the overlap gives the phase shift. We denote the
state of the many body system by a single subscript, so that
$\ket{\phi_i}=\ket{k^{\prime}, \vec{n}}$. The initial state is
$\ket{\phi_0} = \ket{k, \vec{n}_0}$. The $\ket{\phi_0}$ component of
the scattered state after interactions with the gas is given by
\begin{equation}
  S_{0,0} \ket{\phi_0} \approx S_{0,i_{n-1}}^{(N_s)} S_{i_{n-1}, i_{n-2}}^{(N_s-1)} \ldots S^{(1)}_{i_{1},0} \ket{\phi_0}
\end{equation}

Repeated indices are implicitly summed over. The physical process that
corresponds to each set of indices is scattering $\ket{\phi_0}
\rightarrow \ket{\phi_{i_1}} \rightarrow \ket{\phi_{i_2}} \cdots
\rightarrow \ket{\phi_{i_{n-1}}} \rightarrow \ket{\phi_0}$. When any
of these intermediate states is not $\ket{\phi_0}$, we have argued
that the projectile totally decoheres, so the contribution of these
terms to the final coherent state amplitude can be neglected.
\begin{equation}
  \label{eq:S00Approx}
  S_{0,0} \ket{\phi_0} \approx S^{(N_s)}_{0,0} S^{(N_s-1)}_{0,0} \ldots S^{(1)}_{0,0} \ket{\phi_0}.
\end{equation}

The physical interpretation of this expression is that the probability
amplitude for remaining in the initial state is reduced by each
slab. Only this amplitude interferes with the other arm of the
interferometer. $\left| S_{0,0} \right|^2$ is the probability that an
atom in the beam will interfere with itself. The remaining fraction of
the atomic beam contributes only to an incoherent background. The net
result is that the interference fringe contrast is reduced by the
factor $\left| S_{0,0} \right|$. The shift of the interference fringes
compared to a vacuum is given by the argument of $S_{0,0}$.

We will now calculate these quantities by first calculating the
$S^{(j)}_{0,0}$ matrix element due to scattering from a single
slab. Then we may take the product in (\ref{eq:S00Approx}) to obtain $S_{0,0}$.

\section{Calculation of the S-matrix}
The S-matrix is the time evolution operator that takes a quantum state from
the distant past, prior to a collision, into the distant future, after
the collision; that is, $S=\lim_{t \rightarrow \infty} U(t,-t)$. It can be
expressed in terms of the scattering matrix, $T$, as~\cite{Watson1964-2}
\begin{equation}
  S = 1 - 2 \pi i \delta(E - H_o) T,
\end{equation}
where $T$ is defined~\cite{Sakurai1994-2} by
\begin{equation}
  \label{eq:TDef}
  T = V + \lim_{\epsilon \to 0} V \frac{1}{E-H_o+i \epsilon} T. 
\end{equation}
The limit will not appear in what follows; it is understood that we
must take the small $\epsilon$ limit. $V$ and $H_o$ are the $N+1$
particle operators defined above. $S^{(j)}$ is the S-matrix due to the
potential $V^{(j)}$. Replacing $V$ with $V^{(j)}$ in the definition of
$T$ gives $T^{(j)}$. We calculate $S_{0,0}^{(j)}$ by expanding
$S^{(j)} \ket{k, \vec{n}_0}$ to find the coefficient on its $\ket{k,
  \vec{n}_0}$ component,
\begin{equation}
  \label{eq:ScattState}
  S^{(j)} \ket{k, \vec{n}_0}  = \ket{k, \vec{n}_0} - 2 \pi i \delta(E - H_o) T^{(j)} \ket{k, \vec{n}_0}.
\end{equation}
In order to extract the $\ket{k, \vec{n}_0}$ component of the second term
in (\ref{eq:ScattState}), we insert a complete set of
eigenstates of $H_o$ between the delta function and $T^{(j)}$,
\begin{multline}
  \delta(E - H_o) T^{(j)} \ket{k, \vec{n}_0} \\
  = \int \!\! dk^{\prime} \sum_{\vec{n}} \delta \left(
    E-E_{k^{\prime}, \vec{n}} \right) \ket{k^{\prime}, \vec{n}}
  \bra{k^{\prime}, \vec{n}} T^{(j)} \ket{k, \vec{n}_0}.
\end{multline}
The terms of the sum with $\vec{n} \neq \vec{n}_0$ are orthogonal to
$\ket{k, \vec{n}_0}$. They do not contribute to $S_{0,0}^{(j)}$. It is
only necessary to consider the term $\vec{n} = \vec{n}_0$. There, the
argument of the delta function is considerably simplified due to the
cancellation of the energy contribution of the discrete quantum
numbers. In that case, $E - E_{k^{\prime}, \vec{n}_0} = \frac{\hbar^2
  k^2}{2 m} - \frac{\hbar^2 k^{\prime 2}}{2 m}$, where $m$ is the mass
of the projectile. The integral over $k^{\prime}$ can then be
performed easily to find that the coefficient on the $\ket{k,
  \vec{n}_0}$ component of $S^{(j)} \ket{k, \vec{n}_0}$ is
\begin{equation}
  S^{(j)}_{0,0} = 1-i \frac{2 \pi m}{\hbar^2 k} \bra{k, \vec{n}_0} T^{(j)} \ket{k, \vec{n}_0}.
\end{equation}
The net effect of the gas on the amplitude of the initial state is
obtained according to (\ref{eq:S00Approx}) as the product of the
individual slab results,
\begin{equation}
  \label{eq:S00manybody}
  S_{0,0} \approx \prod_{j=1}^{N_s} \left( 1-i \frac{2 \pi m}{\hbar^2 k} \bra{k, \vec{n}_0} T^{(j)} \ket{k, \vec{n}_0} \right).
\end{equation}
$T^{(j)}$ is the full scattering matrix due to a single slab,
including multiple scattering within the slab. In a dilute gas,
sequential scattering from different targets is unlikely within a slab
that is much thinner than the length of the gas cell. Neglecting
multiple scattering \textit{within} single slabs, the $N+1$ particle
matrix element of $T^{(j)}$ reduces to a sum of 2-particle matrix
elements (see Appendix~\ref{sec:t-matrix-approx}),
\begin{equation}
  \label{eq:TjApproximation}
  \bra{k, \vec{n}_0} T^{(j)} \ket{k, \vec{n}_0} \approx \sum_{i=1}^{N} \bra{\chi, \varphi_i} t^{(j)}_{0i} \ket{\chi, \varphi_i},
\end{equation}
where $\chi$ designates the initial state of the projectile,
$\varphi_i$ designates the initial state of the $i^{\text{th}}$
particle, and $t^{(j)}_{0i}$ is the scattering matrix for the
$0^{\text{th}}$ and $i^{\text{th}}$ particles without any other atoms
present,
\begin{equation}
  \label{eq:tjDefinition}
  t^{(j)}_{0i}=V_{0i}^{(j)} + V_{0i}^{(j)} \frac{1}{E_0+E_i-H_0-H_i+i \epsilon} t^{(j)}_{0i}.
\end{equation}

The expression for $S_{0,0}$, excluding multiple scattering within
individual slabs, is
\begin{equation}
  \label{eq:S00Complete}
  S_{0,0} \approx \prod_{j=1}^{N_s} \left( 1-i \frac{2 \pi m}{\hbar^2 k} \sum_{i=1}^{N} \bra{\chi, \varphi_i} t^{(j)}_{0i} \ket{\chi, \varphi_i} \right).
\end{equation}
This result gives the complex probability amplitude for the component
of the projectile that remains coherent after interactions with the
gas. We have explicitly taken into account multiple scattering. It
remains to examine the limit in which the dimensions of the gas cell
and waveguide become arbitrarily large. This will allow us to remove
the artificial confinement depicted in Fig.~\ref{fig:waveguide}. We
find that the result is independent of the confinement and that a
solution of the coherent wave equation emerges directly from these
considerations, without invoking the concept of an average
wavefunction \cite{Lax1951}. Even for an arbitrarily large cell, the
projectile may remain partially coherent after scattering from a
completely free gas. This resolves the conflict between the
experimental results and our expectations based on the usual
expression for scattering in free space.

The initial state of each target and the transverse states of the
projectile appropriate to the waveguide and gas cell are
box-normalized plane waves. Along the $z$ direction, the projectile
remains a free particle. $\ket{\chi, \phi_i}$ in
(\ref{eq:S00Complete}) will be denoted using the wave vectors of the
projectile and target as $\ket{\vec{k_0}, \vec{k_i}}$. The
normalization of $\ket{\vec{k_0}, \vec{k_i}}$ reflects the free nature
of the projectile along the longitudinal axis of the waveguide,
\begin{equation}
  \bracket{\vec{r}_0, \vec{r}_i}{\vec{k}_0, \vec{k}_i} = \frac{e^{i \vec{k}_0 \cdot \vec{r}_0}}{a \sqrt{2\pi}} \frac{e^{i \vec{k}_i \cdot \vec{r}_i}}{a \sqrt{l}}.
\end{equation}
When we convert this expression to center-of-mass coordinates, we must
allocate the normalization constants,
\begin{equation}
  \bracket{\vec{r}_{0i}, \vec{R}_{0i}}{\vec{k}_{0i}, \vec{K}_{0i}} =
  \frac{e^{i \vec{K}_{0i} \cdot \vec{R}_{0i}}}{a \sqrt{2\pi}}
  \frac{e^{i \vec{k}_{0i} \cdot \vec{r}_{0i}}}{a \sqrt{l}}, 
\end{equation}
$\vec{R}_{0i}$ is the center-of-mass coordinate and
$\vec{r}_{0i}=\vec{r}_0 - \vec{r}_i$ is the relative coordinate of the
$0^{\text{th}}$ and $i^{\text{th}}$ particles. The center-of-mass
momentum, $\vec{K}_{0i} = \vec{k}_0 + \vec{k}_i$, is normalized to the
waveguide and the relative momentum, $\vec{k}_{0i} = \frac{m_i}{m +
  m_i} \vec{k}_0 - \frac{m}{m + m_i} \vec{k}_i$, is normalized to the
dimensions of the gas cell. $m$ is the mass of the projectile and
$m_i$ is the mass of the $i^{\text{th}}$ target.

The potential $V^{(j)}_{0i}$ depends only on the relative coordinates
of the projectile and $i^{\text{th}}$ target, with the exception that
it vanishes if the target coordinates lie outside of the
$j^{\text{th}}$ slab. When the range of the potential is much smaller
than the width of the slab, this has the effect of limiting the domain
of the matrix element $\bra{\vec{k}_0, \vec{k}_i} V^{(j)}_{0i}
\ket{\vec{k}_0, \vec{k}_i}$ to the $j^{\text{th}}$ slab. As such, only
when both particles are in the slab is there a contribution to the
matrix element. This requires that the center-of-mass coordinate must
also be in the slab. In principal, the domain of the relative
coordinate that contributes to the matrix element depends on the
position of the center-of-mass coordinate relative to the slab
boundaries, but for local potentials we may take the domain of the
relative coordinate $\vec{r}_{0i}$ to be all space, and replace
$t_{0i}^{(j)}$ with $t_{0i}$. $t_{0i}$ is obtained by replacing
$V_{0i}^{(j)}$ with $V_{0i}$ in (\ref{eq:tjDefinition}). Using these
assumptions, we can rewrite
\begin{equation}
  \bra{\chi, \varphi_i} t^{(j)}_{0i} \ket{\chi, \varphi_i} \approx \bra{\vec{k}_{0i}} t_{0i} \ket{\vec{k}_{0i}} \bracket{\vec{K}_{0i}}{\vec{K}_{0i}}.
\end{equation}
Taken over a slab, $\bracket{\vec{K}_{0i}}{\vec{K}_{0i}} = w/2\pi =
l/(2 \pi N_s)$. Substituting this result into (\ref{eq:S00Complete})
gives,
\begin{equation}
  \label{eq:S00StillSlabs}
  S_{0,0} \approx \prod_{j=1}^{N_s} \left( 1-i \frac{2 \pi m}{\hbar^2 k} \sum_{i=1}^{N} \bra{\vec{k}_{0i}} t_{0i} \ket{\vec{k}_{0i}} \frac{l}{2 \pi N_s} \right) .
\end{equation}
The expression under the product sign in (\ref{eq:S00StillSlabs}) does
not depend on the slab index $j$. We can express the matrix element of
$t_{0i}$ in terms of the forward scattering amplitude in the center of
mass frame, $f(k_{0i}, 0)$, of the $0^{\text{th}}$ and
$i^{\text{th}}$ particles~\cite{Merzbacher} and rewrite the sum over
the particles as $N$ times the average,
\begin{equation}
  S_{0,0} \approx \left( 1+i\, 2 \pi \frac{N}{a^2}
    \frac{1}{N_s}  \left< \frac{f(k_{0i},
        0)}{\mu_{0i} k / m} \right> \right)^{N_s}.
\end{equation}
$N/a^2 = \rho \, l$ is the column density of the gas. We have
explicitly written $\mu_{0i}$ to indicate the reduced mass for each
combination of the projectile and a target. In the case of a target
gas comprised of a single species of atom, we will write $\mu$ for the
reduced mass. The situation in which the projectile velocity dominates
the target velocities permits us to simplify, $k_{0i} \approx (\mu/m)
\, k$~\cite{Dalgarno2001,Dalgarno2002}. As the number of slabs becomes
large, and the width of each slab becomes small compared to the length
of the gas cell, $S_{0,0}$ approaches
\begin{equation}
\label{eq:CWES00}
S_{0,0} \approx \exp \left( i 2 \pi \rho l \left< \frac{f(k_{0i}, 0)}{k_{0i}} \right> \right).
\end{equation}
We may take the dimensions of the waveguide and gas cell to be
arbitrarily large under the condition that we also choose the number
of target atoms so that the column density remains
fixed. Eq.~(\ref{eq:CWES00}) is valid in this free space limit, and
gives precisely a solution of the coherent wave equation when we take
the incident projectile wave function to be a plane wave. 

This central result accounts completely for the phase shift and
persistence of coherence after multiple scattering with a dilute,
many-body quantum mechanical target of free particles. The probability
of remaining in the coherent state decays as $\left| S_{0,0} \right|^2
= e^{-\rho \sigma l}$, where $\sigma$ is the average quantum mechanical
scattering cross section,
\begin{equation}
  \sigma = \left< \frac{4 \pi}{k_{0i}} \Im{ f(k_{0i},
        0)} \right>.
\end{equation}
The cross section is proportional to the imaginary part of the forward
scattering amplitude, whereas the phase shift of the interference
fringes is proportional to the real part,
\begin{equation}
\Delta \phi = \left< \frac{2\pi}{k_{0i}}\, \rho l\, \Re{f(k_{0i},0)} \right>.
\end{equation}
The ratio of the real and imaginary parts of the forward scattering
amplitude, which is directly measured in interferometric measurements
such as~\cite{Pritchard1995-1}, characterizes the extent that the
interference pattern can be shifted before it is washed out due to
decoherence. For weak interactions -- the typical situation in
interferometry -- the real part of the forward scattering amplitude is
proportional to the interaction potential, $V$, whereas the imaginary
part is second order in $V$ (see
Appendix~\ref{sec:phase-shift-vs-attenuation}). Consequently, the
phase shift acquired by the projectile can be made large by increasing
the column density of the gas, while the loss of contrast, 
\begin{equation}
  1 - \left| S_{0,0} \right| \approx 2 \pi \rho l\, 
\Im{\left< \frac{f(k_{0i},0)}{k_{0i}} \right> } \propto V^2,
\end{equation}
remains smaller by a factor of the interaction strength. The
difference in the dependence on the interaction strength clarifies the
ability of a seemingly sensitive, free gas to generate a large phase
shift on a projectile wavefunction due to scattering. This occurs
essentially without loss of contrast if the target gas is sufficiently
dilute and weakly interacting with the projectile.

\section{Low Energy Projectile: Pseudopotential}
\label{sec:pseudopotential}

It is well known that a collection of potential centers, which are
assumed to form a uniform medium in a thin slab, can give rise to an
index of refraction for matter waves~\cite{Fermi1950}. Furthermore,
\cite{Lax1951} has shown that a finite collection of scattering
centers can, when the scattered waves are appropriately averaged, act
as a medium. The scattered wave is the so-called coherent wave, which
suffers attenuation due to the averaging process. We have shown here that
even a finite collection of recoiling quantum mechanical particles in
free space can act as a refractive medium. In addition, decoherence is
a natural consequence of entanglement with the target particles.

In order to illustrate the broader context of our results, it is
instructive to compare the phase shift we obtain for a special case of
the interaction potential with the well-known results of pure
potential scattering. When the projectile is moving slowly relative to
the target atoms, only s-wave scattering needs to be considered, and
we can model the interaction as a contact potential,
\begin{equation}
  V_{0i} = V_o \, \delta( \vec{r}_0 - \vec{r}_i ).
\end{equation}
Recall that the coefficient on the coherent state after interactions
is given by
\begin{equation}
  \label{eq:S00Recollection}
  S_{0,0} \approx \prod_{j=1}^{N_s} \left( 1-2 \pi i \frac{m}{\hbar^2 k} \sum_{i=1}^N \bra{\chi, \phi_i} T_{0i}^{(j)} \ket{\chi, \phi_i} \right).
\end{equation}
For weak potentials, we may approximate $T_{0i}^{(j)}$ to first order
in a Dyson series expansion as
\begin{equation}
  \label{eq:FirstOrderT}
  T_{0i}^{(j)} \approx V_{0i}^{(j)}.
\end{equation}
The matrix element $\bra{\chi, \phi_i} V_{0i}^{(j)} \ket{\chi,
  \phi_i}$ is readily computed using box normalized plane waves as
before,
\begin{equation}
  \label{eq:CPMatrixEl}
  \bra{\chi, \phi_i} V_{0i}^{(j)} \ket{\chi, \phi_i} = \frac{V_o}{2 \pi a^2} \frac{1}{N_s}.
\end{equation}
Substituting (\ref{eq:CPMatrixEl}) into (\ref{eq:S00Recollection})
gives
\begin{equation}
  S_{0,0} \approx \exp \left( i \frac{2 \pi}{k} \left( \frac{-m V_o}{2 \pi \hbar^2} \right) \rho l \right).
\end{equation}
$S_{0,0}$ is a pure phase factor in this approximation. The gas acts
as a medium with an index of refraction for the projectile matter
wave, producing a phase shift \mbox{$\phi = \frac{2 \pi}{k} \left(
    \frac{-m V_o}{2 \pi \hbar^2} \right) \rho l$}.

A direct calculation \cite{Fermi1950} that ignores the quantum state of
the gas atoms and treats them as potential centers leads to the phase
shift
\begin{equation}
  \phi = -\frac{2 \pi}{k} a_o \rho l, 
\end{equation}
where the scattering length $a_o$ can be determined from the solution
to the delta potential scattering problem \cite{Wodkiewicz1991},
\begin{equation}
  a_o = \frac{m V_o/(2\pi \hbar^2)}{1+i k m V_o/(2\pi \hbar^2)} \approx m V_o/(2\pi \hbar^2).
\end{equation}
The approximation of $a_o$ is valid under the same conditions as our
expansion of the T-matrix. The first order term in
(\ref{eq:S00Recollection}) gives precisely the result for the phase
shift that is obtained due to potential scattering. If we were to keep
terms up to second order in the expansion of $T^{(j)}_{0i}$, $S_{0,0}$
would also reduce the amplitude of the coherent state, giving rise to
decoherence.

\section{Conclusions}

We have calculated the effect on a coherent matter wave due to a free
gas environment, in the context of a separated arm atom
interferometer. We have shown that it is possible to discretize the
scattering problem with the artifices of a waveguide and a gas
cell. In this way we have been able to extract the part of the system
that does not become entangled with the environment, but which
nonetheless acquires a phase shift due to a ``recoiless'' interaction
with the free gas. When the free gas is not observed, the state of the
system is obtained by a trace over the environment degrees of freedom,
which causes decoherence. The coherent part of the matter wave
interferes with itself after interacting with the gas, revealing both
the phase shift acquired and the loss of coherence.

The model of a waveguide and gas cell provides a method by which we
have obtained a quantitative result for the phase shift and
decoherence. These quantities properly depend on the column density of
the free gas but not on the dimensions of the confinement. The
dimensions of the waveguide and gas cell may be taken to be
arbitrarily large; therefore, the results are also valid in free
space. We have resolved, therefore, the free space conundrum that
seemed to severely limit the possibility of coherent interactions with
a free gas. Our results show that there is a finite
probability that the atom may scatter from a free gas particle,
acquire a phase shift, but not change the state of the gas particle at
all. The implication is that an interaction took place, but did not
cause the free gas to recoil. The probability of a single target atom
removing a projectile from the coherent beam is the quantum mechanical
scattering cross section.

The projectile must avoid the cross sections of all the gas atoms in
order to remain coherent after passing through the gas sample. We have
shown above that it is possible for a projectile to do so and still
acquire a large phase shift due to the weaker dependence of the cross
section, $O(V^2)$, than the phase shift, $O(V)$, on the strength of
the interaction potential. Physical insight into the dominance of the
phase shift over the scattering cross section in this regime can be
had by realizing, as suggested by~\cite{Pritchard1995-1}, that it is
small impact parameters that contribute to the cross section, and
large impact parameters that contribute to the phase shift. It is
precisely at large impact parameters, when scattering is avoided, that
the phase shift goes as $\delta_l$, the $l^{\text{th}}$ partial wave
phase shift, and the cross section goes as $\delta_l^2$. In this
region, $\delta_l$ is small, and the phase accumulates much faster
than scattering occurs. A projectile that is passing through a dilute
gas will interact at long range with the targets and is operating in
this regime.

The requirement that a projectile evade the scattering cross sections
of the targets as it skirts its way through the free gas becomes an
impossibility for interaction potentials which have long-range
forces. A particularly common example is the Coulomb potential, for
which the total scattering cross section diverges. Such an interaction
potential between the coherent projectile and a target comprised of
free particles should completely suppress interference fringes, even
for a very low density target gas. 

The results we have obtained for the phase shift and the decoherence
are consistent with the multiple scattering treatment of
\cite{Lax1951}, and the coherent state that we extract after
interactions with the environment is a solution of the coherent wave
equation. This very satisfying result shows that a careful analysis, in
which we explicitly incorporate the quantum mechanical state of the
free gas, gives rise to Lax's average wavefunction.

The techniques that we have developed to calculate the effect of a
free gas as a which-way detector explain the surprising robustness of
the spatial coherence in an atom interferometer to interactions with
free particles. The calculations we have done also lay the groundwork
for future investigations into the impact of other, many-body atomic
systems on a coherent atom, due to scattering interactions.

\begin{acknowledgements}
  The authors gratefully acknowledge useful discussions with David
  Pritchard. This work was supported by the National Science
  Foundation through a grant to the Harvard-MIT Center for Ultracold
  Atoms and by the Alexander von Humboldt foundation.
\end{acknowledgements}

\appendix

\section{Approximation of the Scattering T-Matrix}
\label{sec:t-matrix-approx}

We wish to exclude multiple scattering from the initial state diagonal
matrix element of an $N+1$ particle scattering matrix $T$,
\begin{equation}
  \left< T \right> = \bra{\chi, \phi_1, \ldots, \phi_N} T \ket{\chi, \phi_1, \ldots, \phi_N}.
\end{equation}
The Hamiltonian is a sum of operators acting only on the Hilbert
spaces of the indicated particles,
\begin{multline}
  H = H_o+V \\
  = (H_0 + \cdots + H_N) + (V_{0i} + \cdots + V_{0N}).
\end{multline}
Recall the definition~\cite{Sakurai1994-2} of the corresponding T-matrix is
\begin{gather}
  \label{eq:Tdef}
  T = V+VG_oT \\
  G_o = \lim_{\epsilon \rightarrow 0} \frac{1}{E-H_o+i\epsilon}
\end{gather}
We will introduce the operators $T_{01}, \ldots, T_{0N}$ that satisfy
\begin{gather}
  T = \sum_{i=1}^N T_{0i} \\
  \label{eq:T0i}
  T_{0i} = V_{0i} + V_{0i} G_o T_{0i} + \sum_{j\neq i} V_{0i} G_o
  T_{0j}.
\end{gather}
It is only the third term on the right-hand side of (\ref{eq:T0i})
that contributes to multiple scattering. The expression for $T_{0i}$
that excludes multiple scattering is
\begin{equation}
  \label{eq:T0iapprox}
  T_{0i} \approx V_{0i} + V_{0i} G_o T_{0i}.
\end{equation}
This approximation of $T_{0i}$ differs from the definition of the
2-particle scattering matrix, $t_{0i}$, by the replacement of the
$N+1$-particle operator, $G_o$, with a 2-particle operator, $g_{0i}$.
\begin{gather}
  \label{eq:tdef}
  t_{0i} = V_{0i} + V_{0i}\,g_{0i}\,t_{0i},\\
\label{eq:2-body-green}
g_{0i} = \lim_{\epsilon \rightarrow 0} \frac{1}{\left( E_0 + E_i
  \right) - \left( H_0 + H_i \right) + i \epsilon}.
\end{gather}
Consider $i=1$, and note that
\begin{multline}
  \label{eq:Tandt}
  \bra{\phi_2, \ldots, \phi_N} T_{01} \ket{\phi_2, \ldots, \phi_N} \approx \\
  V_{01} + V_{01}\,g_{01} \bra{\phi_2, \ldots, \phi_N} T_{01}
  \ket{\phi_2, \ldots, \phi_N}.
\end{multline}
We have used
\begin{equation}
  \bra{\phi_2, \ldots, \phi_N} G_{o} = \bra{\phi_2, \ldots, \phi_N} g_{01}.
\end{equation}
(\ref{eq:Tandt}) is identical to (\ref{eq:tdef}), so when multiple
scattering is ignored, we can identify
\begin{equation}
  t_{01} \approx \bra{\phi_2, \ldots, \phi_N} T_{01} \ket{\phi_2, \ldots, \phi_N}.
\end{equation}
This result is the same for any $i$. Summing the contributions due to
each $\left< T_{0i} \right>$ gives the approximation we desired
\begin{equation}
  \left< T \right> \approx \sum_{i=1}^N \bra{\chi, \phi_i} t_{0i} \ket{\chi, \phi_i}.
\end{equation}

\section{Expansion of the Scattering Amplitude}
\label{sec:phase-shift-vs-attenuation}

We seek the dependence of the real and imaginary parts of the forward
scattering amplitude, $f(k_{0i}, 0)$, on the interaction potential,
$V_{0i}$, between the projectile and the $i^{th}$ target. This may be
accomplished by relating the scattering amplitude to the two-body
scattering matrix, $t_{0i}$, defined in
(\ref{eq:tdef}),~\cite{Merzbacher}
\begin{equation}
\bra{\vec{k}_{0i}} t_{0i} \ket{\vec{k}_{0i}} = \frac{-2 \pi
  \hbar^2}{\mu a^2 l} f(k_{0i}, 0)
\end{equation}
$f(k_{0i}, 0)$ is the forward scattering amplitude in the center of
mass frame of the projectile and target. $\mu$ is the reduced mass,
$a$ and $l$ are the previously defined dimensions of the gas cell, and
$k_{0i}$ is the relative wavevector. Expanding $t_{0i}$ in a Dyson
series to second order gives
\begin{equation}
\label{eq:t0iexpansion}
t_{0i} \approx V_{0i} + V_{0i} g_{0i} V_{0i}.
\end{equation}
Separating the real and imaginary parts of the two-body Green's
function, $g_{0i}$, gives~\cite{Watson1964-3} 
\begin{multline}
  g_{0i} = \text{P} \frac{1}{(E_0+E_i)-(H_0+H_i)} \\
  - i \pi \delta((E_0+E_i)-(H_0+H_i)).
\end{multline}
Substituting this result into~(\ref{eq:t0iexpansion}), we find that the
real part of the two-body scattering matrix is first order in
$V_{0i}$, whereas the imaginary part is second order,
\begin{align}
\Re{t_{0i}} &\approx V_{0i} \\
\Im{t_{0i}} &\approx -\pi V_{0i} \delta((E_0+E_i)-(H_0+H_i)) V_{0i}.
\end{align}
Therefore, $\Im{f(k_{0i}, 0)}$ is a factor of the interaction
potential smaller than $\Re{f(k_{0i}, 0)}$. 

\bibliography{bibliography}

\end{document}